\documentclass[10pt,conference]{IEEEtran}

\IEEEoverridecommandlockouts
\usepackage[numbers,sort&compress]{natbib}

\makeatletter

\makeatletter

\usepackage{amsmath,amssymb,amsfonts}
\usepackage{algorithmic}
\usepackage{graphicx}
\usepackage{textcomp}
\usepackage{xcolor}
\def\BibTeX{{\rm B\kern-.05em{\sc i\kern-.025em b}\kern-.08em
    T\kern-.1667em\lower.7ex\hbox{E}\kern-.125emX}}

\usepackage{physics}

\usepackage[framemethod=tikz]{mdframed}
\newmdenv[linewidth=0.5pt,
          roundcorner=2pt,
          backgroundcolor=gray!5,
          skipabove=\medskipamount,
          skipbelow=\medskipamount]{qasmcode}

\usepackage[hidelinks]{hyperref}
\usepackage{orcidlink}

\begin{document}

%\title{Sparse and Truncated State Vector Simulation of Peaked Circuits\\
\title{A Sparse and Truncated State Vector Simulator for Peaked Circuits\\
%\title{Approximate Simulation of Peaked Circuits with a Sparse and Truncated State Vector Approach\\
}

\author{\IEEEauthorblockN{Diogo R. Ferreira\orcidlink{0000-0001-5818-9406}}
\IEEEauthorblockA{
\textit{Instituto Superior Técnico (IST)} \\
\textit{Universidade de Lisboa}\\
Lisbon, Portugal \\
\href{mailto:diogo.ferreira@tecnico.ulisboa.pt}{diogo.ferreira@tecnico.ulisboa.pt}}
}

\bstctlcite{BSTcontrol}

\maketitle

\begin{abstract}
In a class of quantum circuits known as peaked circuits, the goal is to predict the most probable bit string at the output of the circuit. Since these circuits are designed to have a sharp peak in their output distribution, in principle it should be possible to simulate them using a truncated state vector with a limited number of terms, or a fraction of the total probability mass. This approximate simulation can be carried out on a classical computer with a sparse representation that stores only the nonzero amplitudes of the state vector, in contrast to the dense representations that are common in most quantum simulators. For efficiency, all operations on the state vector should be vectorized to the furthest possible extent and, if available, hardware acceleration can also be used. This work describes how these requirements were met in an open-source implementation, and discusses its performance and limitations.
\end{abstract}


\section{Introduction}

To set up the context and terminology, we recall some basic concepts about state vectors. For a single-qubit system, the state vector can be written as:
\begin{equation}
	\ket{\psi} = \alpha_0\ket{0} + \alpha_1\ket{1}, \; \text{with} \; |\alpha_0|^{2} + |\alpha_1|^{2} = 1
	\label{eq:psi}
\end{equation}
where $\ket{0}$ and $\ket{1}$ are the basis states, and $\alpha_0$ and $\alpha_1$ are complex amplitudes, whose squared magnitudes (interpreted as probabilities) sum to 1.

A unitary operation on a single qubit can be written as:
\begin{equation}
	U =
	\begin{bmatrix}
		u_{00} & u_{01} \\
		u_{10} & u_{11}
	\end{bmatrix}, \; \text{with} \; U^{\dagger}U=I
	\label{eq:U}
\end{equation}
where $U^{\dagger}$ denotes the adjoint (conjugate transpose) of $U$.

Applying this unitary $U$ on a single qubit yields a new state vector $\ket{\psi'} = \alpha'_0\ket{0} + \alpha'_1\ket{1}$, with amplitudes:
\begin{equation}
	\begin{aligned}
		\alpha'_0 & = u_{00}\alpha_0 + u_{01}\alpha_1 \\
		\alpha'_1 & = u_{10}\alpha_0 + u_{11}\alpha_1 \\
	\end{aligned}
	\label{eq:rule}
\end{equation}

In a multiple-qubit system,
\begin{equation}
	\ket{\psi} = \sum_{i=0}^{2^n-1}\alpha_i\ket{i}, \; \text{with} \; \sum_{i=0}^{2^n-1}|\alpha_i|^{2} = 1
	\label{eq:psi_n}
\end{equation}
where $\ket{i}$ is a basis state of the $n$-qubit system (usually expressed in binary form, in the range $\ket{00...00}$ to $\ket{11...11}$), and $\alpha_i$ is the corresponding amplitude.

In an $n$-qubit system, applying $U$ on a single qubit involves applying the expressions in Eq.~(\ref{eq:rule}) to every pair of amplitudes $(\alpha_{i}, \alpha_{j})$ whose basis states differ only in that qubit (i.e. $\ket{i}$ has $0$ at the corresponding bit position, and $\ket{j}$ has $1$) while having the same bit pattern in all other qubits.

Therefore, in an $n$-qubit system, an operation on a single qubit might require updating all amplitudes across an entire state vector with $2^n$ terms. The same applies to operations on two, three or more qubits, where the unitary matrix has dimensions $4\!\times\!4$, $8\!\times\!8$ or in general $2^m\!\times\!2^m$ for $m\!\leq\!n$ qubits.

In practice, it often happens that many amplitudes are zero, which avoids the need to handle a full-blown state vector with $2^n$ terms. For example, in most quantum circuits, the convention is that the circuit begins in the ground state (where $\alpha_0\!=\!1$, and $\alpha_i\!=\!0$ for $i\!>\!0$), so the initial state of the system can be represented with a single term. However, the number of nonzero terms can grow rapidly:
\begin{itemize}

	\item A unitary operation that creates superposition (e.g. the Hadamard gate) may double the number of nonzero terms in the state vector, because a qubit that was in the ground state (with an amplitude for $0$) now has an amplitude for $0$ and an amplitude for $1$. All fixed patterns for other qubits will have to be duplicated and their amplitudes updated.
	
	\item For an operation that creates entanglement between two qubits, the number of nonzero terms in the state vector could quadruple (although the CNOT gate, for example, operates on two qubits and may create entanglement, but it just permutes the amplitudes without increasing the number of nonzero terms).

	\item Operations on three and four qubits could potentially multiply the number of nonzero terms by 8 and 16, but in practice such gates typically rearrange the amplitudes without increasing the number of terms (as is the case with the three-qubit and four-qubit Toffoli gates).
		
\end{itemize}

In summary, the application of circuit operations may cause the number of terms to grow exponentially from the ground state. However, the fact that there is a limited set of gates with specific forms (see \cite{cross17openqasm} for a list of standard gates) suggests that, at least during an initial stage of the circuit, the number of terms may grow in a manageable way.

As soon as the number of terms reaches the limit of what is possible to store in a classical computer, exact simulation becomes out of reach, but it might still be possible to proceed with an approximate simulation if the state vector is truncated to its most significant terms. Such truncation, as we propose here, can be based on a hard limit (e.g. the maximum number of terms that can be stored in the available memory) or on a threshold (e.g. a minimum fraction of probability mass that should be kept during simulation).

In this work, we describe a sparse and truncated state vector approach, whose main goal is to find the most probable output bit string in peaked circuits~\cite{aaronson24peaked}. These circuits are designed so that their output distribution is peaked on a specific bit string, which occurs with higher probability than the rest. Structurally, peaked circuits are built by composing random-looking circuit layers with peaking layers that selectively boost the amplitude of a target bit string, without making the overall circuit easily distinguishable from a generic random circuit.

This paper is organized as follows. Section~\ref{sec:sparse} reviews dense and sparse representations for the state vector. Section~\ref{sec:truncated} presents the proposed truncation methods. Section~\ref{sec:vectorized} explains how state-vector evolution and truncation can be implemented in a vectorized way. Section~\ref{sec:accelerated} discusses how those operations can be hardware-accelerated on GPUs. Section~\ref{sec:application} describes the application to peaked circuits and analyzes performance scaling. Finally, Section~\ref{sec:conclusion} summarizes the main contributions and directions for future work.

\section{Sparse State Vectors}
\label{sec:sparse}

Most quantum circuit simulators store the full state vector as an array of length $2^n$, and update it in place with gate-specific kernels~\cite{mcguffin05tutorial}. For $n$ qubits, the simulator keeps $\ket{\psi}$ as a flat array of $2^n$ complex numbers, and each gate is implemented as a small linear transformation applied to slices or strides of this array (e.g. $2\!\times\!2$ for one qubit, $4\!\times\!4$ for two-qubit gates, and so on). This maximizes locality and processor utilization, but uses $O(2^n)$ memory, which limits practical simulations to 25–30 qubits on a typical machine.

Examples of circuit simulators using a dense representation include Qiskit~\cite{javadi24quiskit}, Qulacs~\cite{suzuki2021qulacs}, Cirq~\cite{developers25cirq} and PennyLane~\cite{bergholm22pennylane}. On the other hand, there has been a growing interest in sparse simulators, as demonstrated in recent works such as GraFeyn~\cite{westrick24grafeyn}, qblaze~\cite{venev25qblaze} and SparQSim~\cite{sun2026sparqsim}. The relevance of sparse simulators is further demonstrated by the fact that the default simulator in Microsoft QDK (Quantum Development Kit, an extension to the popular Visual Studio Code editor) leverages a sparse representation~\cite{jaques2022sparsity}.

Perhaps the most impressive of these works is qblaze~\cite{venev25qblaze}, which introduces a cache-friendly sparse representation together with carefully designed multi-core algorithms that scale efficiently to multiple CPUs, yielding significant speedups over prior sparse tools and mainstream dense simulators.

In any case, sparse tools such as qblaze, despite using a sparse representation, are still \emph{exact} state-vector simulators, i.e. they keep track of all nonzero amplitudes that arise during state evolution. For this reason, they are advantageous when the state vector stays sparse and there are multi-core CPUs to exploit. When state vectors become fully or mostly dense, a dense simulator will usually perform better.

\section{Truncated State Vectors}
\label{sec:truncated}

In addition to making use of a sparse representation, in this work we look at how the state vector can be pruned or truncated in order to keep a manageable number of terms. By \emph{manageable}, we mean a number of nonzero amplitudes that fits the resource limits (in terms of memory and processing capabilities) of a classical computer.

Naturally, by truncating the state vector, we are abandoning the prospect of carrying out exact simulation. This decision should not be taken lightly, when there have been significant efforts and progress in circuit optimization~\cite{karuppasamy25review}, and these advances are making it easier to perform exact simulation of increasingly large circuits on classical computers~\cite{liu25simulability}.

On the other hand, even though classical simulation keeps improving, a truncated state-vector approach addresses an important gap, by allowing a trade-off between fidelity and significant savings in memory and time, especially for highly entangled circuits, where the exponential growth of the state vector is still a fundamental obstacle.

A truncated state-vector approach addresses this problem by retaining only a \emph{relevant} portion of the state (e.g. the basis states carrying most of the probability mass), while discarding a controlled tail and then renormalizing. This plays a similar role to bond-dimension truncation in tensor-network simulations, where small Schmidt coefficients are dropped to keep the representation tractable~\cite{vanhecke25tangent}.

Here, the main motivation is that in some application scenarios (e.g. peaked output distributions, variational or heuristic algorithms, noisy hardware), the observables of interest are dominated by a small subset of basis states, so a carefully designed truncation scheme can yield orders‑of‑magnitude savings in memory and runtime, while maintaining accuracy within experimental or algorithmic tolerances.

A recent paper~\cite{miller26approximate} proposes a way to simulate quantum circuits by keeping only a sparse subset of basis states at every step. The state is represented as a list of $k$ basis indices and amplitudes, and a bitwise contraction algorithm applies two-qubit gates directly to this sparse list. After each gate, the list may temporarily grow, but is truncated back to size $k$ by discarding terms, primarily using a top-$k$ rule that keeps the largest-magnitude amplitudes and renormalizes.

An interesting feature of that work is that it analyzes how truncation affects fidelity, showing that the average fidelity is well approximated by the total probability mass retained after truncation. For this reason, in this work we propose two truncation methods:
\begin{itemize}

	\item A hard limit on the number of terms (i.e. a maximum number of nonzero amplitudes to keep, together with their basis states), which we refer to as top-$k$ truncation.
	
	\item A threshold on probability mass (i.e. a minimum fraction of probability mass to keep, based on the sum of squared magnitudes), which we refer to as $p$-mass truncation.

\end{itemize}

Both methods keep the largest-magnitude amplitudes (and thus the fewest number of terms) to satisfy the chosen limit or threshold. When both limit and threshold are specified, $p$-mass truncation is applied first and, if the resulting number of terms still exceeds the limit, top-$k$ truncation is applied last. In any case, the state vector is renormalized after truncation.

Bounding the number of terms gives a direct handle on runtime and memory. On the other hand, what matters for fidelity is how much probability mass is retained, so a criterion on probability mass is more natural to control the accuracy of observables. By offering both top-$k$ truncation and $p$-mass truncation, one can budget either for resources or for accuracy, and then use the other method as a secondary knob to navigate the trade-off between efficiency and fidelity.

\section{Vectorized Operations}
\label{sec:vectorized}

In practice, a sparse and truncated state-vector simulator will be spending most of its time applying the same few algebraic operations to many amplitudes, basis indices, and array elements. Expressing these updates in terms of vectorized operations allows the underlying numerical libraries to take advantage of cache locality, special processor instructions, and multi‑threaded routines.

Vectorization is especially important in our setting, where the simulator repeatedly updates amplitudes and applies truncation. Even if the number of terms remains small compared to $2^n$, the array of amplitudes may still contain millions of elements. By formulating the evolution and truncation of the state vector as bulk array transformations, the implementation can better leverage modern CPU and GPU architectures.

\subsection{State-Vector Evolution}

To illustrate how operations can be vectorized, we start by looking at how a unitary operation updates the state vector:
\begin{itemize}

	\item In the expressions of Eq.~(\ref{eq:rule}), $\alpha_0$ is multiplied by the first column of $U$, and $\alpha_1$ is multiplied by the second column of $U$. The same rule applies when the unitary operates on a single qubit in an $n$-qubit system. In this case, for every pair of amplitudes $(\alpha_i, \alpha_j)$ whose basis states differ only in that qubit, $\alpha_i$ is multiplied by the first column of $U$, and $\alpha_j$ is multiplied by the second column of $U$.
	
	\item For an operation on two qubits, instead of a pair of amplitudes, we need to consider every tuple of four amplitudes $(\alpha_{i_{00}}, \alpha_{i_{01}}, \alpha_{i_{10}}, \alpha_{i_{11}})$ whose basis states have the corresponding bit pattern in those two qubits, and differ only in those bits. In this case, the unitary has dimensions $4\!\times\!4$, and each of its four columns is multiplied by the corresponding amplitude in the 4-tuple.

	\item In general, for a unitary operation on $m\!\leq\!n$ qubits, we need to consider every tuple of $2^m$ amplitudes for which the basis states have a different pattern in those $m$ qubits and an equal pattern in the remaining qubits. In this case, the unitary operator has dimensions $2^m\!\times\!2^m$, and each of its $2^m$ columns is multiplied by the corresponding amplitude in the $2^m$-tuple.

\end{itemize}

This reasoning allows to identify which column of $U$ should be multiplied by each amplitude in the state vector. Once those columns have been identified, the unitary operation can be applied as a single multiplication across all amplitudes at once. 

In a second step, the new basis states can be elicited from the bit patterns that correspond to the nonzero amplitudes. In Eq.~(3), we see that $\alpha_1$ contributes to $\alpha'_0$, and $\alpha_0$ contributes to $\alpha'_1$. In general, any amplitude $\alpha_i$ may contribute to a new amplitude $\alpha'_j$. If the old amplitude $\alpha_j$ was zero, but $\alpha'_j$ is not, then there is a new nonzero amplitude, with a corresponding basis state $\ket{j}$. Since $\ket{j}$ is usually expressed in binary form, it corresponds to the bit pattern of $\alpha'_j$. Therefore, the new basis states can be obtained from the new amplitudes through bitwise operations in a vectorized way.

Finally, we need to handle collisions. In Eq.~(3), we see that both $\alpha_0$ and $\alpha_1$ contribute to $\alpha'_0$ (and the same applies to $\alpha'_1$). In general, for a unitary operation on $m$ bits, each new amplitude might be the result of $2^m$ contributions (if all those contributions are nonzero). Typically, $m$ is small, since most circuit gates typically operate on one, two, or three qubits at most. Nevertheless, it is necessary to collect the $2^m$ contributions and add them up. 

When multiple contributions to the same basis state need to be added up, it becomes necessary to group them by basis state and sum within each group. In a vectorized implementation, this takes the form of a \emph{segmented sum}~\cite{sobczyk25segmented}. Although the segmented sum is particularly challenging to implement on parallel hardware, it is now well supported in modern CPU and GPU libraries. Still, this step is often the main performance bottleneck during state evolution.

\subsection{State-Vector Truncation}

Regarding state-vector truncation, whether based on top-$k$ or $p$-mass, the array operations are more easily parallelizable. In this case, the main bottleneck is the sorting of terms from largest to smallest probability (i.e., by decreasing order of the squared magnitude of amplitudes). Once the amplitudes are sorted, we take the first terms up to the specified limit ($k$), or whose cumulative probability ($p$) satisfies the desired threshold. A final step renormalizes the state vector.

\section{Hardware Acceleration}
\label{sec:accelerated}

To leverage hardware acceleration, we developed a GPU backend that replicates the vectorized operations on the CPU with equivalent operations on the GPU. Since the implementation was already formulated in a vectorized manner, this substitution required only minor code changes. This approach avoids the need for hand-written kernels or architecture-specific tuning, and keeps the CPU and GPU backend code nearly identical, simplifying maintenance.

For numerical precision, both backends use the same data types: 128‑bit complex numbers for amplitudes, 64‑bit floating point values for their real and imaginary parts, and 64‑bit integers for basis states. Using 64-bit integers imposes a theoretical limit of $2^{64}$ basis states and an upper bound of $n\!\leq\!64$ qubits. In practice, this constraint is not an issue, since the available memory can be exhausted well before that, even with a sparse and truncated representation.

With the GPU backend, the state-vector arrays (amplitudes and basis states) remain on the GPU device throughout the entire simulation. Only the final results (such as the most probable bit string) are transferred back to the host. Preliminary benchmarks indicate a speedup of about an order of magnitude when switching from the CPU to the GPU backend, although memory, in this device, can be more constrained.

\section{Application to Peaked Circuits}
\label{sec:application}

Peaked circuits~\cite{aaronson24peaked} constitute an important class of quantum circuits whose output probability distributions are dominated by a small number of basis states. In such cases, the amplitudes are strongly concentrated around one or a few peaks, while the remaining states contribute negligibly to the overall probability mass. This structure makes them particularly suitable for simulation with a sparse and truncated state vector, since only the dominant basis states need to be represented.

In theory, this idea should work, and indeed an important theoretical result has been established for \emph{peaked shallow circuits}~\cite{bravyi24shallow}. These are circuits where the outcome of each output qubit depends only on a constant-size neighborhood of input qubits, regardless of the total number of qubits in the system. For these circuits, it has been shown that their output distribution can be approximated by a classical algorithm with runtime $n^{O(\log n)}$, which is consistent with the existence of sparse descriptions involving a quasi-polynomial number of terms, instead of the full $2^n$ dimensional state.

In practice, the scenario can be far more challenging. When circuits become very deep or strongly entangling, it is possible to have a peaked output distribution and yet have a nontrivial fraction of the probability mass spread over an enormous number of basis states, which are not fully accounted for during state evolution. To keep the truncation error under control, it becomes necessary to retain far more amplitudes, and the number of relevant basis states can grow from quasi-polynomial in $n$ towards a substantial fraction of $2^n$. In other words, the same peak height no longer implies that only a small subset of basis states is relevant.

This means that the proposed approach may perform very unevenly across different peaked circuits. In some cases, when most of the probability mass is consistently concentrated on a relatively small set of basis states, the method can successfully identify the most probable output despite truncation of the state vector. In other cases, when the circuit produces a single prominent peak but still distributes a significant portion of the probability mass over an enormous number of basis states, the same truncation strategy can fail dramatically.

Since the approach may succeed in some instances while failing on others, its behavior should be investigated on a variety of peaked circuits. For this purpose, the peaked-circuit hackathons organized by BlueQubit~\cite{gharibyan25peaked} provide an ideal playground and an interesting set of examples. One of such examples is a circuit known as \emph{sharp peak}, which is built from layers of arbitrary single‑qubit rotations (u3), interleaved with controlled-Z (cz) entangling gates between neighboring qubits along the chain, as illustrated in Fig.~\ref{fig:qasm}.

\begin{figure}[t]
	\centering
	\begin{qasmcode}
		\ttfamily\footnotesize
		OPENQASM 2.0;\\
		include "qelib1.inc";\\
		qreg q[44];\\
		...\\
		u3(2.425927,0,-pi/2) q[0];\\
		u3(0.770504,0,pi/2) q[1];\\
		cz q[0],q[1];\\
		...\\
		u3(2.595146,0,-pi/2) q[1];\\
		u3(1.401835,0,pi/2) q[2];\\
		cz q[1],q[2];\\
		...\\
		u3(2.856619,0,-pi/2) q[42];\\
		u3(1.401101,0,pi/2) q[43];\\
		cz q[42],q[43];\\
		...\\
		u3(2.097514,0,-pi/2) q[0];\\
		u3(1.377591,0,pi/2) q[43];\\
		cz q[0],q[43];\\
		...
	\end{qasmcode}
	\vspace{-\baselineskip}
	\caption{Excerpt of BlueQubit's \emph{sharp peak} circuit.}
	\label{fig:qasm}
\end{figure}

In addition to the cz-gates between neighboring qubits (i.e. 0--1, 1--2, ..., 42--43), there is a wrap-around coupling 0--43, which connects all qubits in a ring structure. This ring connectivity across 44 qubits and 580 instructions yields a large, deep, and highly entangling circuit, which is unfavorable for dense simulators and also for tensor network methods such as MPS (matrix product states), because accurately representing the state requires large bond dimensions~\cite{cirac21matrix}.

\subsection{Circuit Simulation Strategy}

To mitigate the cost of simulating peaked circuits, such as the one in Fig.~\ref{fig:qasm}, we employ a sparsity-aware gate-reordering and gate-fusion strategy:
\begin{itemize}

	\item First, we reorder the gates so that, during simulation, the set of currently active qubits is kept as small as possible, while the introduction of additional qubits is postponed for as long as the circuit dependencies allow. The goal is to contain (or delay) the exponential growth of nonzero terms in the state vector, while preserving the logical sequence of the original circuit (because we only reorder commuting gates or dependency-compatible gates, in a similar spirit to lazy qubit-reordering strategies for parallel simulators~\cite{teranishi25lazy}).

	\item Then, for each block of single- and two-qubit gates (as in the blocks depicted in Fig.~\ref{fig:qasm}), we fuse those gates into a multi‑qubit unitary operation. This gate fusion strategy can significantly reduce the number of updates to the state vector (from one update per gate to one update per block), and it has been used successfully to accelerate classical simulation in other recent works as well~\cite{kawase26fusion,kumaresan26gpu}. In our case, gate fusion also makes truncation less aggressive, since it is applied once at the end of each block rather than after every individual gate.
	
\end{itemize}

In summary, circuit simulation begins by reordering the gates and identifying the blocks. It then proceeds by iterating over those blocks; for each block, it updates the state vector based on a fused unitary, followed by truncation according to the chosen method (top-$k$ or $p$-mass).

\subsection{Performance Scaling}

With this simulation strategy, it is possible to find the correct output bit string for the sharp peak circuit using a very small number of terms ($<\!2^5$). However, other circuits may require substantially more terms in order to correctly identify the most probable bit string. In that case, it is important to understand how the simulation performance is expected to scale as the size of the state vector is allowed to grow.

\begin{figure}[ht]
	\centering
	\includegraphics[width=\linewidth]{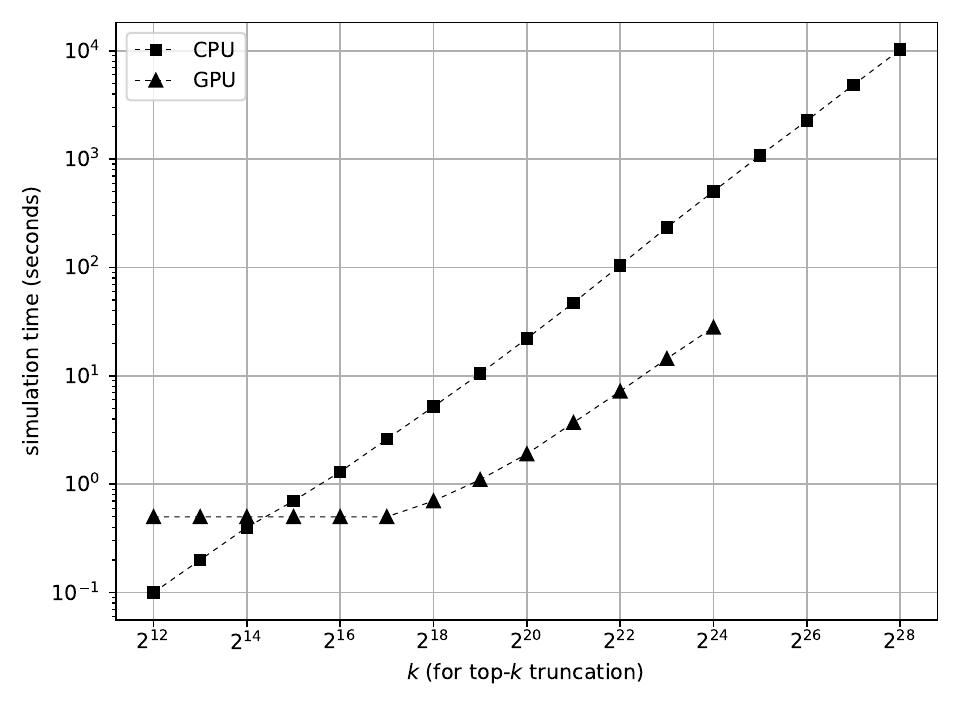}
	\vspace{-\baselineskip}
	\caption{Circuit simulation time when applying top-$k$ truncation.}
	\label{fig:plot_1}
\end{figure}

For this purpose, we use top-$k$ truncation to analyze how simulation time increases with $k$. Fig.~\ref{fig:plot_1} shows that the relationship is linear across multiple orders of magnitude, i.e. the simulation time is proportional to $k$. This is particularly evident for the CPU version; with the GPU backend, there is a fixed overhead that is apparent at low values of $k$. On the other hand, for larger values of $k$, the GPU version is about an order of magnitude faster, up to the point where the device memory does not allow a further increase in $k$.

\begin{figure}[ht]
	\centering
	\includegraphics[width=\linewidth]{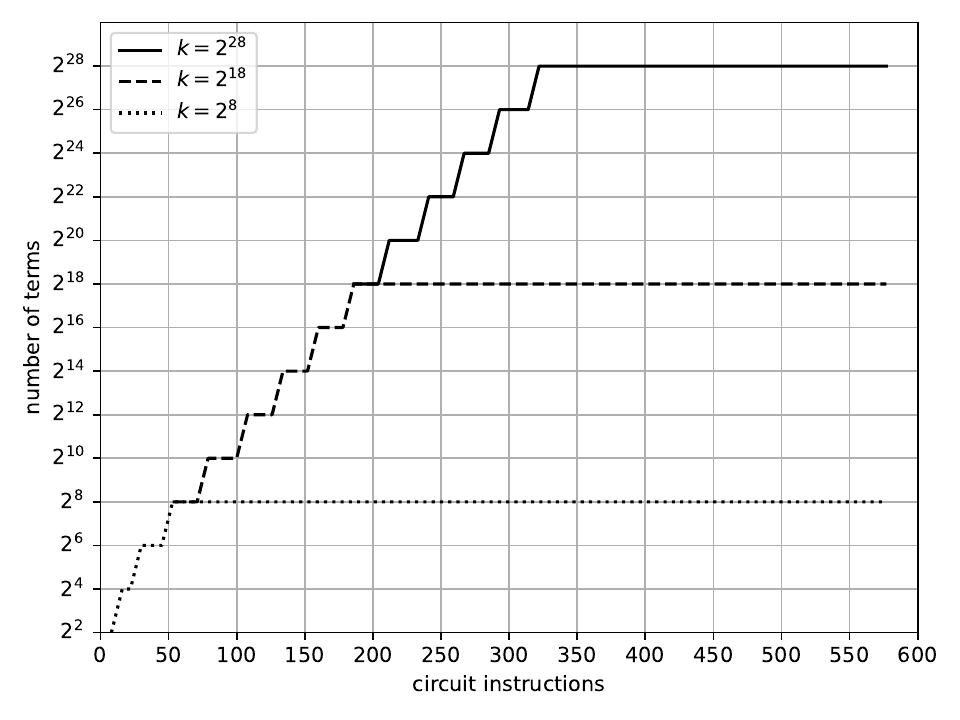}
	\vspace{-\baselineskip}
	\caption{Number of terms during circuit simulation with top-$k$ truncation.}
	\label{fig:plot_2}
\end{figure}

Fig.~\ref{fig:plot_2} illustrates how the number of terms increases during circuit simulation, for three different values of $k$. In an initial stage, some blocks of circuit instructions do not change the size of the state vector, while others cause the number of terms to double, so the staircase pattern in the plot is actually an exponential growth, up to the limit established by $k$. Naturally, the larger the number of terms becomes, the longer it will take to process each block of circuit instructions.

\begin{figure}[ht]
	\centering
	\includegraphics[width=\linewidth]{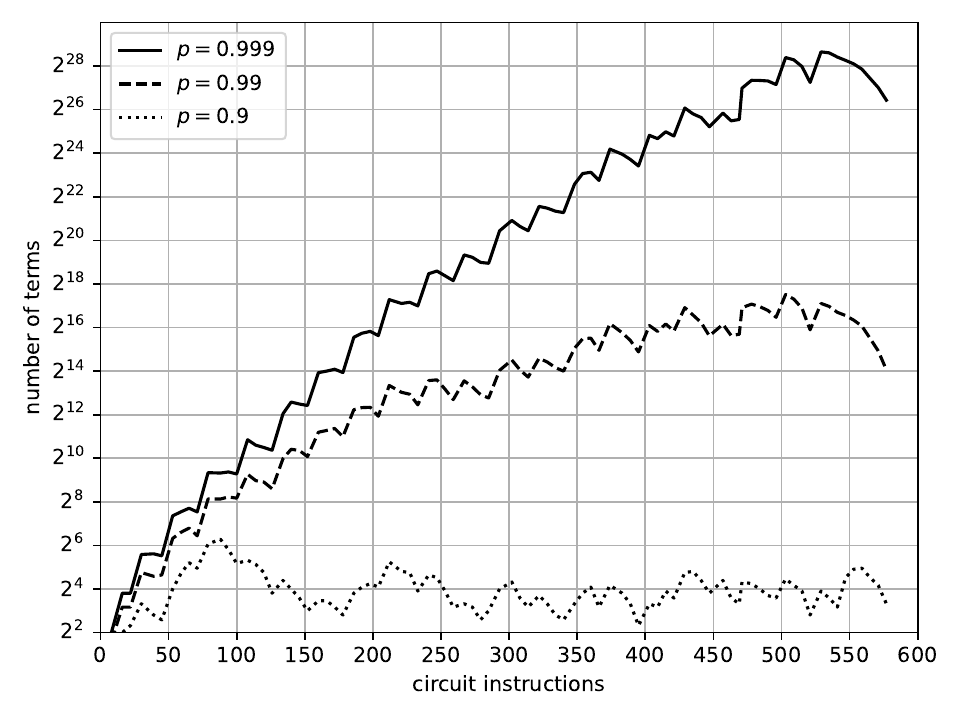}
	\vspace{-\baselineskip}
	\caption{Number of terms during circuit simulation with $p$-mass truncation.}
	\label{fig:plot_3}
\end{figure}

With $p$-mass truncation, the situation is different. Here, the state vector is truncated after each block (as opposed to top-$k$ truncation, where it is allowed to grow freely up to a point). However, if the specific fraction of probability mass is high, the state vector may grow well above any anticipated limit. Fig.~\ref{fig:plot_3} illustrates this situation, where a $p$-mass fraction of 99.9\% allows the number of terms to grow past $2^{28}$. On the other hand, lower fractions of probability mass keep the number of terms fairly contained.

\begin{figure}[ht]
	\centering
	\includegraphics[width=\linewidth]{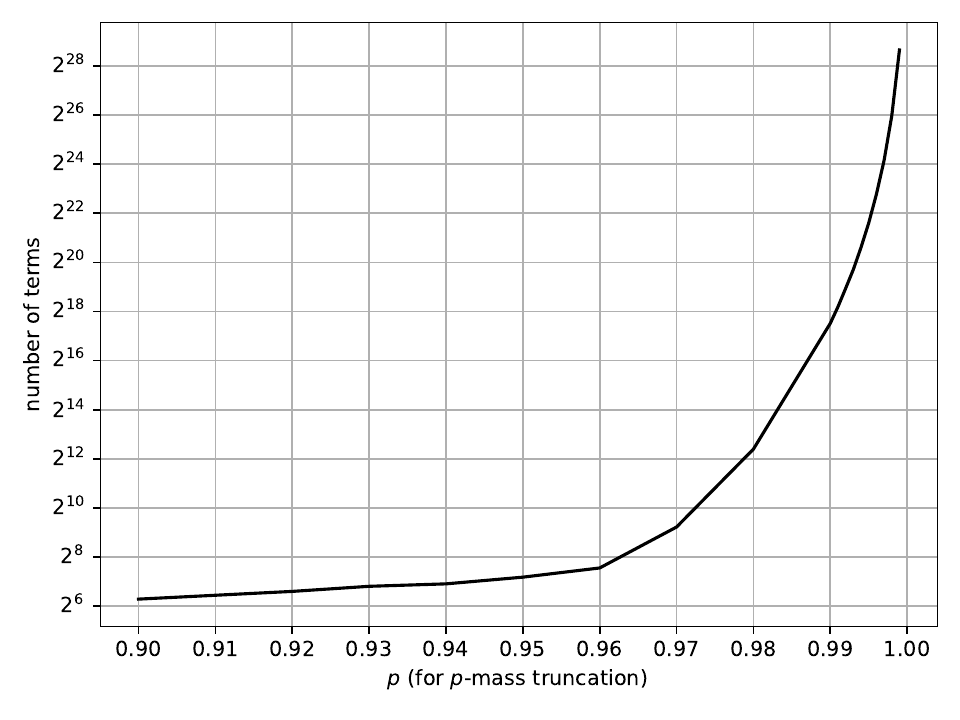}
	\vspace{-\baselineskip}
	\caption{Number of terms vs. $p$-mass threshold.}
	\label{fig:plot_4}
\end{figure}

Finally, Fig.~\ref{fig:plot_4} shows that, as the fraction of probability mass approaches 100\%, the number of terms observed during circuit simulation grows very rapidly towards $2^n$, where $n$ is the number of qubits. At the same time, the steepness of this region suggests that, in practice, it may be possible to identify the most probable bit string of a peaked circuit using a number of terms that remains well below $2^n$, while still keeping a high fraction of the probability mass.

\section{Conclusion}
\label{sec:conclusion}

In this work, we introduced a sparse and truncated state-vector simulator, with vectorized operations implemented efficiently for both CPU and GPU backends. Since our goal is to apply this simulator to peaked circuits, we also devised a circuit simulation strategy based on gate reordering and gate fusion, with the aim of delaying the exponential growth of the number of terms in the state vector.

When such growth becomes untenable, the main approach to contain it is through two truncation methods: one imposes a hard limit on the number of terms, while the other retains a fraction of the total probability mass. Both methods prioritize the largest-amplitude components, under the assumption that these will suffice to identify the most probable output of a peaked circuit, provided that the probability mass remains concentrated on a few basis states.

Although the approach works for the peaked circuit considered here and a few others, a carefully drafted circuit that spreads the probability mass across a very large number of basis states can undermine any truncation method. Despite this limitation, we believe that the proposed approach is a useful addition to the toolbox of circuit simulators, complementing existing exact and approximate techniques.

For more difficult peaked circuits, preprocessing techniques such as ZX-based optimization~\cite{fischbach26zx} and other graph-based methods~\cite{lee26coloring} can be very helpful. In future work, we plan to integrate these techniques into our circuit simulation strategy. Since these optimizations act prior to the actual simulation, they may contribute to further enhance the effectiveness of the sparse and truncated simulator presented here.

\section*{Acknowledgments}

Some of the experiments described in this work were performed using a GPU donated by NVIDIA Corporation. The author is also grateful for the motivating discussions with Jorge Domingues at Johnson \& Johnson MedTech.

%The author is affiliated with a research unit that is financially supported by FCT (\emph{Fundação para a Ciência e Tecnologia, I.P.}) through grants UID/50010/2025, UID/PRR/50010/2025, UID/PRR2/50010/2025, and LA/P/0061/2020.

\section*{Data Availability}

The code and data supporting this work are available in the repository: \url{https://github.com/diogoff/qstvec/}

\newpage

\bibliographystyle{IEEEtran}
\bibliography{paper}

\end{document}